\documentclass[%
 preprint,
 amsmath,amssymb,
 aps,
]{revtex4-1}

\usepackage{graphicx}
\usepackage{dcolumn}
\usepackage{bm}
\usepackage{xcolor}
\usepackage{tikz}
\usetikzlibrary{quantikz}
\usepackage[version=4]{mhchem}
\usepackage{braket}
\usepackage{soul}
\usepackage{hyperref}

\begin{document}
\title{Simulating the electronic structure of spin defects on quantum computers}

\author{Benchen Huang}
\affiliation{Department of Chemistry, University of Chicago, Chicago, IL 60637, USA}%


\author{Marco Govoni}
 \email{mgovoni@anl.gov}
 \affiliation{Pritzker School of Molecular Engineering, University of Chicago, Chicago, IL 60637, USA}
\affiliation{Materials Science Division and Center for Molecular Engineering, Argonne National Laboratory, Lemont, IL 60439, USA}%

\author{Giulia Galli}
\email{gagalli@uchicago.edu}
\affiliation{Department of Chemistry, University of Chicago, Chicago, IL 60637, USA}
\affiliation{Pritzker School of Molecular Engineering, University of Chicago, Chicago, IL 60637, USA}
\affiliation{Materials Science Division and Center for Molecular Engineering, Argonne National Laboratory, Lemont, IL 60439, USA}%

\date{\today}

\begin{abstract}
We present calculations of both the ground and excited state energies of spin defects in solids carried out on a quantum computer, using a hybrid classical/quantum protocol. We focus on the negatively charged nitrogen vacancy center in diamond and on the double vacancy in 4H-SiC, which are of interest for the realization of quantum technologies. We employ a recently developed first-principle quantum embedding theory to describe point defects embedded in a periodic crystal, and to derive an effective Hamiltonian, which is then transformed to a qubit Hamiltonian by means of a parity transformation. We use the variational quantum eigensolver (VQE) and quantum subspace expansion methods to obtain the ground and excited states of spin qubits, respectively, and we propose a promising strategy for noise mitigation. We show that by combining zero-noise extrapolation techniques and constraints on electron occupation to overcome the unphysical state problem of the VQE algorithm, one can obtain reasonably accurate results on near-term-noisy architectures for ground and excited state properties of spin defects.
\end{abstract}

\maketitle


\section{Introduction} \label{introduction}

Quantum simulations of the physical and chemical properties of molecules and solids ~\cite{jones2015, krylov2018, schleder2019, maurer_advances_2019, bogojeski_quantum_2020} are crucial  to gain insight into a wide range of complex problems, for example catalytic reactions~\cite{bell_quantum_2011, xu_theoretical_2019, hammes2021integration} and the search for optimal materials for sustainable energy sources~\cite{lee2021impact}  and quantum technologies~\cite{weber2010, wolfowicz2021}. One of the essential ingredients of quantum simulations is the solution of the electronic structure problem for molecules and solids, namely the time-independent Schr\"odinger equation of interacting electrons in the field of atomic nuclei. Such solution provides the basis for the evaluation of numerous ground and excited state properties of matter. However, the algorithms used at present on classical computers to solve the electronic structure problem, especially those based on wavefunction methods, face serious  computational bottlenecks  due to the  exponential growth of the dimension of the many-body wavefunction  as a function of system size~\cite{cao2019}.

Quantum computers hold promise to drastically improve our ability to carry out quantum simulations of many-electron systems by taking advantage of superposition and entanglement principles offered at the hardware level by quantum bits (qubits)~\cite{cao2019, head2020quantum, bauer2020}. $N$ qubits can represent $2^N$ complex numbers, which would require $2^{N+7}$ bits to be represented in double precision on classical computers, and some problems, such as the solution of the Schr\"odinger equation, may benefit from the memory scaling. Whether one can achieve quantum advantage in solving useful chemistry and physics problems on quantum computers is still under debate. However, efforts to develop algorithms to simulate molecules and solids on quantum computers~\cite{aspuru-guzik2005,omalley2016,shen2017,kandala2017,santagati2018,hempel2018,ryabinkin2018,mccaskey2019,li2019,smart2019,sagastizabal2019,google2020,metcalf2020,nam2020,gao2021,rossmannek2021,kawashima2021,teplukhin2021computing,kirsopp2021quantum,jones2021chemistry, li2011,cruz2020,kivlichan2020,montanaro2020,uvarov2020,motta2020,mei2020,bassman2021constant,mineh2021solving,powers2021exploring,cerasoli2020,sureshbabu2021,choudhary2021,yamamoto2021quantum} have been flourishing in the last decade, and several interesting results on ground and excited states of small molecular systems (containing up to a dozen atoms) have appeared in the literature~\cite{aspuru-guzik2005,omalley2016,shen2017,kandala2017,santagati2018,hempel2018,ryabinkin2018,mccaskey2019,li2019,smart2019,sagastizabal2019,google2020,metcalf2020,nam2020,gao2021,rossmannek2021,kawashima2021,teplukhin2021computing,kirsopp2021quantum,jones2021chemistry}.

The number of degrees of freedom and hence the number of atoms  that can be handled at present on near intermediate scale quantum (NISQ) computers is limited, due to the availability of hardware architectures with only a small number of low-fidelity qubits (a 127-qubit device was recently  announced~\cite{ballfirst} but devices used for most calculations appeared in the literature have few tens of qubits). In particular, the hardware limitation poses a challenge for quantum simulations of heterogeneous solids which require the use of supercells with hundreds of atoms. Recent studies have focused on two fronts: (i) reducing the complexity of the simulations of condensed phases by using an effective Hamiltonian to represent a fragment (or active part) of a solid~\cite{ma2020a,mineh2021solving}, thus effectively reducing the number of degrees of freedom; and (ii) developing techniques to mitigate the noise present in NISQ hardware~\cite{endo2021hybrid} which affects the results of calculations on quantum computers. These techniques are based on different methods, including zero-noise extrapolation (ZNE) \cite{li2017efficient, temme2017error,endo2018practical,lloyd2018quantum}, symmetry verification \cite{sagastizabal2019experimental}, quasi-probability methods \cite{temme2017error, endo2018practical}, or stochastic error mitigation \cite{sun2021mitigating}. An alternative strategy to complexity reduction is the use of model Hamiltonians, e.g., Hubbard, Heisenberg~\cite{li2011,cruz2020,kivlichan2020,montanaro2020,uvarov2020,motta2020, mei2020,powers2021exploring,bassman2021constant,mineh2021solving} or tight-binding  Hamiltonians~\cite{cerasoli2020,sureshbabu2021,choudhary2021}.


Recently, we proposed a computational framework~\cite{sheng2021quantum} to carry out the calculation of the electronic structure of an active site in a periodic system using a quantum embedding theory which we call here quantum {\it defect} embedding theory (QDET)~\cite{ma2020a, ma2021} that is suitable, for example, for the study of spin-defects. Spin-defects in semiconductors and insulators are promising candidates for the realization of quantum technologies~\cite{weber2010quantum,pfaff2014unconditional,hsieh2019imaging,wolfowicz2021quantum}, including quantum sensing and communication. The electronic states of spin-defects usually exhibit a multi-reference nature, which requires methods beyond mean-field theories for a proper description.

In this work, we present calculations of both ground and excited states of two spin-defects, i.e., the NV$^-$ center in diamond and the double vacancy (VV) in 4H-SiC described by QDET using a hybrid classical/quantum protocol on a real quantum computer. To the best of our knowledge, the calculations of excited state are reported here for the first time. In addition, we present the application of a correction scheme to impose physical constraints on the output of quantum simulations, and we propose an extrapolation strategy to carry out error mitigation. 

The rest of the paper is organized as follows: in section \ref{Methods} we present the classical and quantum algorithms used in this work and in section \ref{Results} the results of our calculations, including error analysis and mitigation techniques. Section \ref{Conclusions}  concludes the paper with a summary and outlook.

\clearpage

\section{Methods} \label{Methods}

\begin{figure}[h!]
    \centering
    \includegraphics[width=0.5\textwidth]{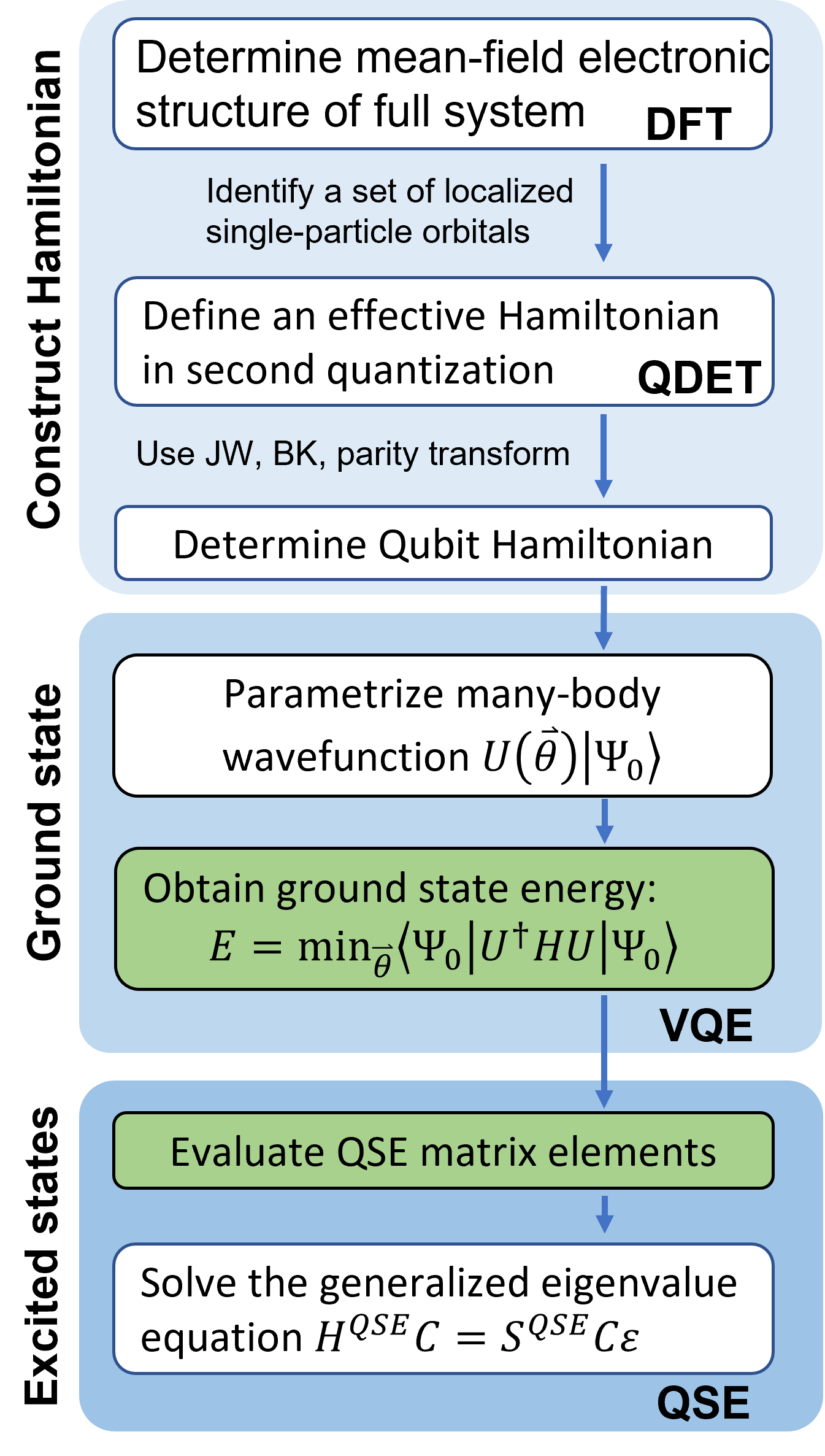}
    \caption{Workflow used to simulate the ground and excited state energies of spin defects, with operations executed on a quantum computer indicated in green. The transformation from a second quantized to a qubit Hamiltonian may be obtained with a Jordan-Wigner (JW), Bravyi-Kitaev (BK) or parity transformation. DFT and QDET denotes calculations carried out using Density Functional Theory and the quantum defect embedding theory, respectively. VQE and QSE denotes the variational quantum eigensolver and quantum subspace expansion algorithms used for ground and excited state calculations, respectively. See text for definition of the equations.}
    \label{fig:summary}
\end{figure}

As mentioned in the introduction, there are numerous problems in condensed matter physics and chemistry, including point defects in semiconductors, that can be naturally formulated in terms of active regions surrounded by a host medium. These problems can be addressed using embedding theories~\cite{libisch2014embedded,wouters2016practical, hermes2019multiconfigurational,pham2019periodic,ma2020,ma2020a,ma2021,lan2017generalized,zgid2017finite} where the electronic structure of the host and the active region are described at different levels of theory, in particular a higher level of theory is chosen for the active region, which can describe the multi-reference character of wavefunctions. The QDET proposed in Ref~\cite{ma2020,ma2020a,ma2021} is an example of embedding theories formulated in terms in Green's functions, and it has been shown to accurately  describe the low-lying excitations of several spin-defects in insulators~\cite{ma2020a}. The computational strategy of our work is centered on QDET and on  calculations of the electronic structure of spin defects on quantum computers. Fig.~\ref{fig:summary} summarizes the methods adopted in this work to carry out mean-field calculations for a chosen supercell, followed by QDET calculations for a defect, and the definition of an effective Hamiltonian for the active space of the defect. The calculation of the ground and excited states of the effective Hamiltonian are carried out on a quantum computer with a variational quantum eigensolver (VQE) and the quantum subspace expansion (QSE), respectively. These methods, starting with QDET, are briefly summarized below.

\subsection{Quantum Defect Embedding Theory to obtain effective Hamiltonians}

We first define a periodically repeated arrangement of several hundreds of atoms (or supercell) representing the point-defect of interest within a given solid, and we compute the electronic structure of the supercell from first principles with Density Functional Theory (DFT) or hybrid-DFT. We then select a sub-set of single particle wave-functions which are localized around the defect and physically represent its electronic states. This sub-set defines an active space whose excitations are described by an effective Hamiltonian $H_{eff}$. The effective potential entering $H_{eff}$ is evaluated by computing the effect of the environment onto the active space with many-body perturbation theory techniques; two-body interactions are evaluated using constrained DFT either within the constrained random-phase approximation (cRPA)~\cite{aryasetiawan2004,aryasetiawan2009,miyake2008} or by including explicitly exchange-correlation effects~\cite{ma2019,nguyen2019}. The effective Hamiltonian includes correlation effects between the electronic states of the active space. In this work we use QDET to describe strongly correlated electronic states of point-defects that are not properly described by single-determinant wavefunctions, and hence by DFT. 

In essence, using QDET one can reduce the complexity of evaluating many-body states of a small guest region embedded in a large host system: the problem is reduced to diagonalizing  a many-body Hamiltonian simply defined on an active space, where the number of degrees of freedom is smaller than that required to describe the entire supercell of hundreds of atoms.

The active spaces of the systems discussed in this work contain less than ten electrons, and  $H_{eff}$ can be solved using the full configuration interaction (FCI)~\cite{helgaker2014} method on a classical computer. However FCI scales exponentially with system size and it has so far been limited to active spaces with up to 22 electrons and 22 orbitals~\cite{vogiatzis2017pushing}, although results of adaptive sampling CI have been reported for 52 electrons and 52 orbitals~\cite{levine2020casscf}. Hence, in order to solve  more complex problems, the opportunity offered by quantum computers appears worth exploring, as the overall scaling of FCI may eventually be overcome with quantum architectures.

\subsection{Variational Quantum Eigensolver to obtain ground state energies} \label{VQE}

As mentioned in the introduction, quantum computers have an exponential memory advantage over classical hardware, which in principle may be harnessed  to compute the eigenstates of a Fermionic Hamiltonian representing a many-body system of electrons. For example, the quantum phase estimation (QPE)~\cite{aspuru-guzik2005} algorithm has been proven to be exponentially faster in finding eigenvalues of unitary operators than any available algorithm  on a classical computer~\cite{kitaev1995quantum, abrams1997simulation, abrams1999quantum}. However, calculations using QPE are still impractical in the absence of  error correction and  require quantum resources that exceed the current capability of NISQ hardware~\cite{elfving2020}.
An alternative algorithm to QPE is the variational quantum eigensolver (VQE)~\cite{peruzzo2014}, where the properties of many-body states are measured on a quantum device, but the parameters that define such states are stored on a classical computer. Therefore, VQE allows one to use shallower circuits than QPE and hence to perform calculations on  non fault-tolerant quantum computers ~\cite{cerezo2021variational}.

In the last decade VQE has been successfully applied to several quantum chemistry problems~\cite{peruzzo2014,omalley2016,shen2017,kandala2017,hempel2018,google2020,metcalf2020, nam2020,gao2021,fedorov2021ab,seetharam2021digital,lee2021simulation,verdon2019quantum,guo2021thermal,powers2021exploring}, including the calculation of the total energies of small molecules~\cite{peruzzo2014,omalley2016,shen2017,kandala2017,hempel2018,google2020,metcalf2020, nam2020,gao2021}, the evaluation of forces by finite differences from Hartree-Fock total energies for the H$_2$ molecule~\cite{fedorov2021ab}, the calculations of the zero-field NMR spectrum of the methyl group of acetonitrile on a trapped-ion quantum computer~\cite{seetharam2021digital}, and the calculations of Heisenberg spin systems using a generalization of VQE for thermal states~\cite{verdon2019quantum}. 

When using VQE, the ground state of the Hamiltonian is approximated by a normalized trial state $\ket{\Psi(\vec{\theta})} = U(\vec{\theta})\ket{\Psi_0}$, where the unitary operator $U(\Vec{\theta})$ is constructed using a set of classical parameters $\vec{\theta} = (\theta_1, \dots, \theta_n)$ and $\ket{\Psi_0}$ is a (usually unentangled) initial state. The expectation value $\braket{E(\vec{\theta})} = \bra{\Psi(\vec{\theta})} \hat{H} \ket{\Psi(\vec{\theta})}$ provides an upper bound to the ground state energy of the system. The energy expectation value is optimized on a classical computer by variationally optimizing the parameters $\vec{\theta}$, with all inputs (energy values) evaluated on a quantum computer. Here we used VQE to find the ground state energy and corresponding eigenvector of effective Hamiltonians obtained using the QDET.

The physical Hamiltonian ($H_{eff}$) is mapped onto a qubit Hamiltonian $\hat{H}_q$, e.g., by using the Jordan-Wigner~\cite{jordan1934}, Bravyi-Kitaev~\cite{bravyi2002}, or parity~\cite{bravyi2017tapering} mapping: $\hat{H}_q = \sum_i g_i \hat{P}_i$, where $g_i$ are coefficients determined by one and two-electron integrals, and $\hat{P}_i = \{I, X, Y, Z\}^{\otimes N}$ are Pauli correlators acting on $N$ qubits. In general, the mapping from $H_{eff}$ to $\hat{H}_q$ does not insure that the Hilbert space of the qubit Hamiltonian is the same as that of the original Hamiltonian and hence it is necessary to impose constraints on the many-body wavefunction through an ansatz to avoid introducing “unphysical states” in the space of $\hat{H}_q$, i.e., states that are not present in the Hilbert space of $H_{eff}$. Furthermore, a proper ansatz should satisfy additional, multiple requirements. It is of course important to choose variational parameters spanning a manifold of states that can accurately approximate the ground state of the system. In addition, the chosen unitary operator $U(\Vec{\theta})$ should be constructed in such a way that it can be implemented with the current capacity of gates and qubit connectivity of NISQ computers. There are two classes of ans\"atze explored in the literature: hardware efficient ones, designed specifically by taking into account the hardware characteristics~\cite{kandala2017}, and chemistry-inspired ones. The former class usually enables the use of short depth quantum circuits at the expense of including a large number of variational parameters, while chemical inspired ones attempt to minimize the number of variational parameters, leading to the need for deeper circuits. 

In this work, we chose the Unitary Coupled Cluster Singles and Doubles (UCCSD) ansatz~\cite{peruzzo2014, romero2018strategies}, which belongs to the chemically inspired class and originates from coupled cluster theory \cite{vcivzek1966correlation, bartlett2007coupled,helgaker2014}. The UCCSD ansatz involves the definition of a unitary operator through the exponential of a couple cluster operator that contains pertinent single and double electronic excitations:
\begin{equation}
    \ket{\Psi} = e^{\hat{T} - \hat{T}^{\dagger}} \ket{\Psi_0},\; \hat{T} = \sum_{i\in \mathcal{A}, a\in \mathcal{V}} \theta_i^a \hat{a}^{\dagger}_a \hat{a}_i + \frac{1}{4}\sum_{i,j\in \mathcal{A};\; a,b\in \mathcal{V}}\theta_{i,j}^{a,b} \hat{a}^{\dagger}_a \hat{a}^{\dagger}_b \hat{a}_j \hat{a}_i,
    \label{eq:UCCSD}
\end{equation}
where $i,j$ are occupied single particle orbitals (belonging to subspace $\mathcal{A}$), $a,b$ are virtual orbitals (belonging to subspace $\mathcal{V}$), $\hat{a}^{\dagger}, \hat{a}$ are the fermionic creation and annihilation operators, and $\theta_i^a, \theta_{i,j}^{a,b}$ are variational parameters. A straightforward implementation of Eq.~\ref{eq:UCCSD} is challenging as the operators describing electronic excitations may not commute. A numerically manageable solution may be obtained by introducing a so called "trotterized" version of the unitary operator defining the UCC ansatz. We consider two electronic excitation operators $\hat{A}$ and $\hat{B}$ that do not commute and the  Trotter-Suzuki formula:
\begin{equation}
    e^{\hat{A}+\hat{B}} = \lim_{n \to \infty} \left(e^{\frac{\hat{A}}{n}} e^{\frac{\hat{B}}{n}}\right)^n. \label{eq:Trotter}
\end{equation}
The evaluation of the right-hand side of Eq.~\ref{eq:Trotter} requires large values of $n$, which is expected not to be practical for NISQ computers as it would require long circuits. An approximation widely adopted in the literature is to use $n = 1$ i.e., $e^{\hat{A}+\hat{B}} \approx e^{\hat{A}} e^{\hat{B}}$ (one-step first-order Trotter-Suzuki approximation). Such an approximation introduces an error (Trotter error) and we notice that the ordering of $A$ and $B$ on the right-hand side could have an impact on this error, as discussed in Ref.~\cite{grimsley2019trotterized}.

\subsection{Quantum Subspace Expansion to obtain excited state energies}

We now turn to the discussion of the calculations of excited states, for which different algorithms are required. The excited states of a fermionic Hamiltonian may be computed on a quantum computer starting from the output of VQE optimizations by using, e.g., the variational quantum deflation (VQD) \cite{higgott2019} or quantum subspace expansion (QSE) algorithms. The former is a constrained variational approach where a penalty term $\langle \Psi_i | \Psi \rangle$ is added to the energy expectation value to enforce orthogonality among states. A prerequisite for using VQD is that the target excited states should be included in the manifold of states spanned by the variational wavefunction defined by the chosen ansatz; such a requirement is not straightforward to insure. The QSE is an alternative strategy to VQD and can be viewed as a quantum analog of FCI. Once a reference wavefunction $\ket{\Psi}$ is prepared, a set of expansion operators $\{\hat{O}_i\}$ is chosen, which act on $\ket{\Psi}$ to form a basis given by $\{\hat{O}_i \ket{\Psi}\}$, where $\hat{O} \in \{\hat{a}^{\dagger}_a \hat{a}_i,\; \hat{a}^{\dagger}_a \hat{a}^{\dagger}_b \hat{a}_j \hat{a}_i | i,j\in\mathcal{A};\; a,b \in \mathcal{V}\}$. We evaluate the Hamiltonian and overlap matrix elements using such basis:
\begin{equation}
    H^{\text{QSE}}_{ij} = \bra{\Psi} \hat{O}^{\dagger}_i \hat{H} \hat{O}_j \ket{\Psi},\;\; S^{\text{QSE}}_{ij} = \bra{\Psi} \hat{O}^{\dagger}_i \hat{O}_j \ket{\Psi}.
\end{equation}
Using the matrices defined above, we then solve the generalized eigenvalue problem in the well conditioned subspace given by
\begin{equation}
    H^{\text{QSE}} C = S^{\text{QSE}} C \varepsilon,
\end{equation}
where $C$ is the matrix of eigenvectors and $\varepsilon$ the diagonal matrix of eigenvalues. As mentioned in section~\ref{VQE}, the Fermion-to-qubit mapping is used to transform the Fermion operators $\hat{a}_i^{\dagger}, \hat{a}_j$ into Pauli operators acting on $N$ qubits, and the matrix elements are evaluated as weighted sums of the expectation values of Pauli correlators. The reference $\ket{\Psi}$ is usually taken to be the ground state. The effectiveness of the QSE relies on a careful choice of creation and annihilation operators which should enable the description of the desired excitations.

In our  work, we use QSE to compute excited states for two reasons: (i) the construction of the Hamiltonian $H^{QSE}$ and overlap matrix $S^{QSE}$ elements may be carried out using the same quantum circuit as those used for VQE~\cite{mcclean2020, mcardle2020} calculations of the ground state, and (ii) the procedure does not require additional quantum resources, but only additional measurements~\cite{colless2018}.

\clearpage

\section{Results} \label{Results}

In this section we describe the results for the many-body ground and excited states of the NV$^-$ center in diamond and VV in 4H-SiC obtained using the IBM Qiskit package~\cite{aleksandrowicz2019qiskit} on the \textit{ibmq\_casablanca} quantum computer. In Appendix~\ref{SI: Hardware} we report the details of the measurement error mitigation procedure~\cite{dewes2012, maciejewski2020mitigation}, applied to all the measurements carried out on a quantum computer.

\subsection{Reference results on classical hardware}

\begin{figure}[h!]
    \centering
    \includegraphics[width=0.7\textwidth]{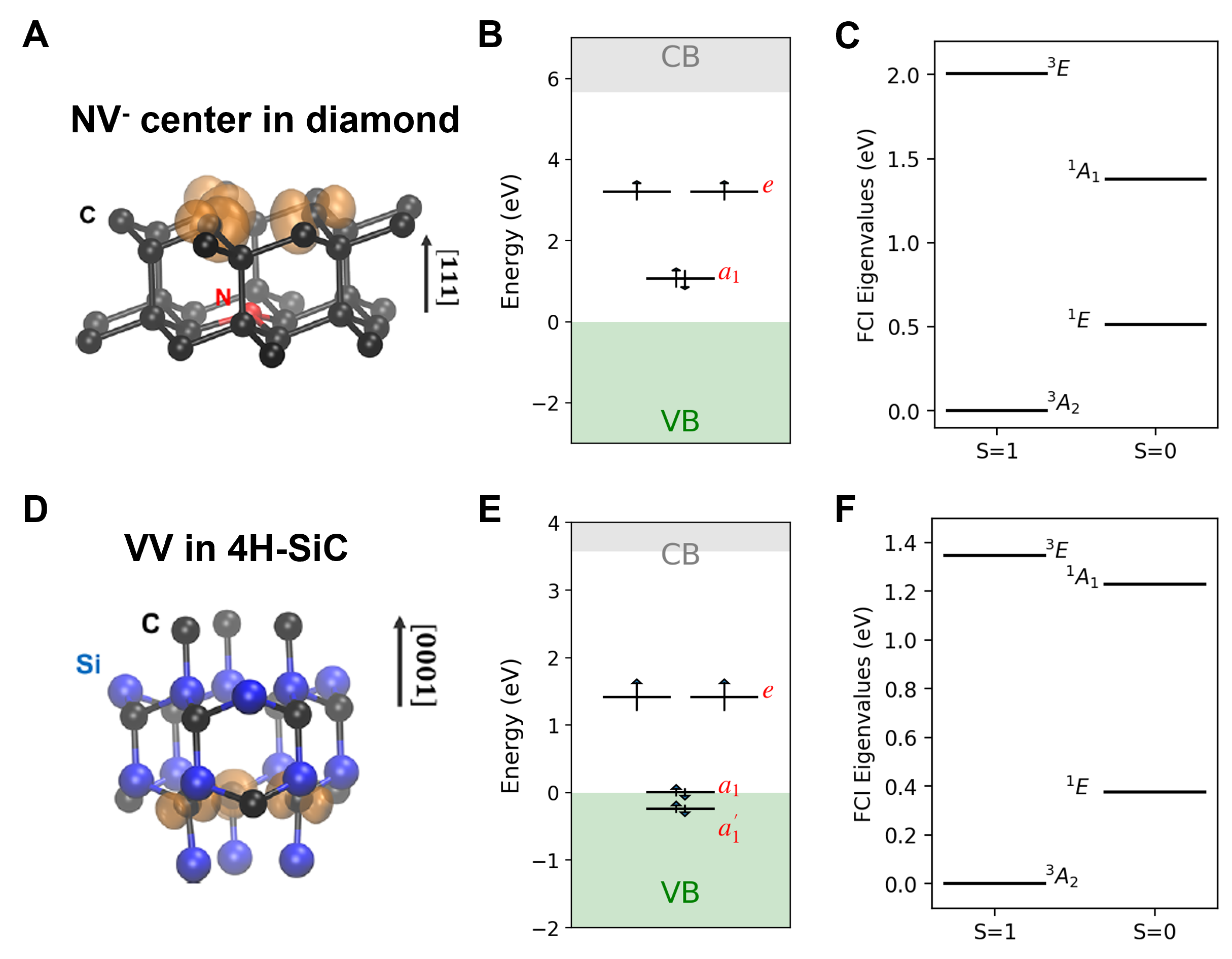}
    \caption{Spin defects studied in this work: the NV$^-$ center in diamond and the VV in 4H-SiC. Panels A and D show a ball-and-stick representation of the defects, where orange iso-surfaces are total spin densities. Panels B and E show single particle states obtained by solving the Kohn-Sham equations for the entire periodic solid, where gray and green shaded areas represent the conduction (CB) and valence band (VB), respectively; the single particles states shown as black lines were used to build the (4e, 3o) and the (6e, 4o) minimum models for the active spaces of the NV$^-$ and VV centers, respectively. Panels C and F show the low-lying many-body energy levels obtained by solving the effective Hamiltonians using the FCI method on classical hardware.}
    \label{fig:classical_preparation}
\end{figure}

The single particle electronic structure of the NV$^-$ center in diamond and the VV in 4H-SiC is computed with hybrid DFT using  216-atom and 200-atom periodic supercells, respectively. We  use  the plane-wave pseudopotential method as implemented in the Quantum Espresso code~\cite{giannozzi2009quantum}, dielectric-dependent hybrid functionals (DDH)~\cite{skone2014}, and a kinetic energy cutoff of 50 Ry. For both systems the active space contains the single particle states localized at the defect site and we refer to the use of this active space as the ``minimum model'', which at present represents the best compromise between accuracy and efficiency (see Fig.~\ref{fig:classical_preparation}). We adopt the QDET embedding scheme as implemented in the WEST~\cite{govoni2015large} code and we evaluate the dielectric screening beyond the random-phase approximation by coupling the WEST and Qbox~\cite{gygi2008architecture} codes, as in Ref.~\cite{ma2021,govoni2021}. The effective Hamiltonian is diagonalized with FCI using the PySCF code~\cite{sun2018pyscf} on a classical hardware and the eigenvalues obtained in this way are considered as reference results for our calculations on a quantum computer. The atomic and electronic structures of the defects studied here, chosen active spaces, and FCI results are summarized in Fig.~\ref{fig:classical_preparation}.

\subsection{Calculation of the ground state using a quantum computer}

The ground state of the effective Hamiltonian constructed for both the NV$^-$ center in diamond and the VV in 4H-SiC is a ${}^3A_2$ triplet state.  Using the minimum models described above (see Fig.~\ref{fig:classical_preparation}), the $m_s=0$ component may be written as a linear superposition of two Slater determinants~\cite{ma2020a}:
\begin{equation}
    \ket{{}^3A_2, m_s=0; \text{NV}} = \frac{1}{\sqrt{2}}\left(\ket{a_1 \bar{a}_1 e_x \Bar{e}_y} + \ket{a_1 \bar{a}_1 \Bar{e}_x e_y} \right),
    \label{eq:3A2_NV}
\end{equation}
\begin{equation}
    \ket{{}^3A_2, m_s=0; \text{VV}} = \frac{1}{\sqrt{2}}\left(\ket{a^\prime_1 \bar{a}^\prime_1 a_1 \bar{a}_1 e_x \Bar{e}_y} + \ket{a^\prime_1 \bar{a}^\prime_1 a_1 \bar{a}_1 \Bar{e}_x e_y} \right),
    \label{eq:3A2_VV}
\end{equation}
for the NV$^-$ and the VV center, respectively. In Eq.s~\ref{eq:3A2_NV}-\ref{eq:3A2_VV}, $a^\prime_1$, $a_1$, $e_x$, $e_y$ (spin-up) and $\bar{a}^\prime_1$, $\bar{a}_1$, $\bar{e}_x$, $\bar{e}_y$ (spin-down) denote the single particle orbitals.

We first focus on the NV$^-$ center and consider two initial states that are used as trial wavefunctions for the VQE algorithm: (a) $\ket{a_1 \bar{a}_1e_x \Bar{e}_x}$, or (b) $\ket{a_1 \bar{a}_1 e_x \Bar{e}_y}$. The UCCSD expression of Eq.~\ref{eq:UCCSD} is used as an ansatz for the unitary operator applied to the trial state. In case (a) the ansatz leads to a trial wavefunction with six variational parameters, whereas in case (b) the trail wavefunction contains only $\theta_{e_x \bar{e}_y}^{e_y \bar{e}_x}$ as variational parameter (for convenience of notation, we refer to this parameter as $\theta$ since all additional parameters may be set to zero to enforce from the start a triplet 
solution). The effective Hamiltonian of both systems is transformed onto a qubit representation using the parity mapping~\cite{bravyi2017tapering}.

\begin{figure}[h!]
    \centering
    \includegraphics[width=0.7\textwidth]{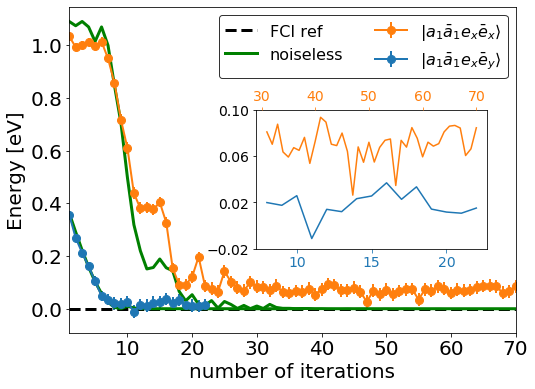}
    \caption{Optimization of the $\ket{{}^3A_2, m_s=0}$ state of the NV$^-$ center using the variational quantum eigensolver (VQE) on \textit{ibmq\_casablanca} and on a noiseless simulator using four qubits. The strongly-correlated $\frac{1}{\sqrt{2}}(|a_1 \bar{a}_1 e_x\bar{e}_y\rangle + |a_1 \bar{a}_1 \bar{e}_x e_y\rangle)$ state is obtained starting from: the $|a_1 \bar{a}_1 e_x\bar{e}_x\rangle$ state, or the $|a_1 \bar{a}_1 e_x\bar{e}_y\rangle$ state. We used the parity transformation to obtain the qubit Hamiltonian acting on four qubits; the optimization was carried out with the COBYLA algorithm~\cite{powell1994direct}. The noiseless simulation was performed with the QASM simulator~\cite{cross2017open}. The zero of energy is the result obtained on classical hardware with the full configuration interaction (FCI) method. In the inset we compare the converged energies obtained from the two chosen trial states.}
    \label{fig:initial_state}
\end{figure}

\begin{figure}[h!]
    \centering
    \begin{quantikz}[row sep=0.5cm, column sep=0.5cm]%
    q_0 & \gate[4]{\text{I}} & \qw & \qw & \qw & \qw & \qw & \gate[4]{\text{M}} & \qw\\
    q_1 & & \gate{H} & \ctrl{2} & \qw & \ctrl{2} & \gate{H} & & \qw\\
    q_2 & & \qw & \qw & \qw & \qw & \qw & & \qw \\
    q_3 & & \gate{R_x(-\frac{\pi}{2})} & \targ{} & \gate{R_z(\theta)} &\targ{} & \gate{R_x(\frac{\pi}{2})} & & \qw
\end{quantikz}
    \caption{Quantum circuit executed on four qubits ($q_0$ to $q_3$): I, and M represent the state initialization and measurement blocks, respectively. The measurement block includes Pauli correlators so as to enable the measurement of observables, e.g., the energy and the electron number. The symbol  $H$ represents a Hadamard gate; $R_{x,z}(\theta)$ represents the rotation of the variational parameter $\theta$ (see text) around the $x,z$ axis.}
    \label{fig:ansatz_circuit}
\end{figure}
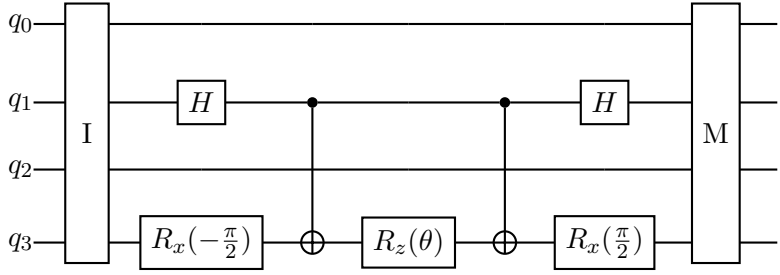

Fig.~\ref{fig:initial_state} shows the convergence of the ground state energy as a function of the number of VQE iterations, starting from either trial states. We obtain convergence to the same results as obtained with FCI when using a quantum simulator (i.e., in the absence of noise) for both trial states; however, we obtain two different values when using \textit{ibmq\_casablanca}, due to the presence of hardware noise. Not surprisingly the impact of noise on the results depends on the choice of the initial state. In case (a), where there are six variational parameters and the circuit depth is 26, we obtain  a $\sim$0.07 eV error, while in case (b), where there is only one variational parameter and the circuit depth is 6, the error is smaller ($\sim$0.02 eV). For case (b), i.e., when the initial state $\ket{\Psi_0}=\ket{a_1 \bar{a}_1 e_x \Bar{e}_y}$, the variational wavefunction in the UCCSD ansatz reads (see Appendix~\ref{SI: Ansatz on trial states}): 
\begin{equation}
    \ket{\Psi(\theta)}=\cos\left(\frac{\theta}{2}\right)\ket{a_1 \bar{a}_1 e_x \Bar{e}_y}+\sin\left(\frac{\theta}{2}\right)\ket{a_1 \bar{a}_1 \Bar{e}_x e_y}. \label{eq:Ansatz}
\end{equation} 
As expected, the energy has a minimum when $\theta=\frac{\pi}{2}$, i.e., for $\ket{\Psi(\frac{\pi}{2})}=\ket{{}^3A_2, m_s=0; \text{NV}}$. In the following, we choose the initial state as in (b) above, and use a simultaneous perturbation stochastic approximation (SPSA)~\cite{spall1992multivariate} optimizer, which has been shown to be robust to noise~\cite{kandala2017}. The corresponding quantum circuit is depicted in Fig.~\ref{fig:ansatz_circuit}. We choose physical qubits such that the mapping leads to a qubit configuration minimizing faulty CNOT gates (see Appendix~\ref{SI: Hardware}).

\begin{figure}[h!]
\includegraphics[width=0.6\textwidth]{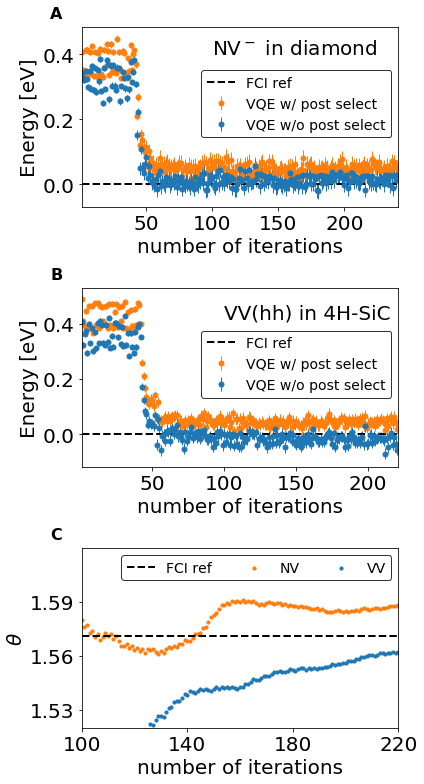}
\caption{Optimization of the ground state energy of the $\text{NV}^-$ in diamond A and VV(hh) in 4H-SiC B carried out with the variational quantum eigensolver (VQE) algorithm using four and six qubits respectively on \textit{ibmq\_casablanca}, with (orange dots) and without (blue dots) post-selection of states (see text). The full configuration interaction (FCI) energy is reported for reference. In panel C we show the variation of the parameter $\theta$ (see Eq.~\ref{eq:Ansatz}) in the VQE optimization; the value is obtained from averaging the parameter at the end of the VQE optimization. The dashed line corresponds to the exact solution, i.e., $\theta=\frac{\pi}{2}$.}
\label{fig:results}
\end{figure}

The energy obtained with VQE as a function of the number of iterations is reported in Fig.~\ref{fig:results} A and B (blue dots). We find that several measurements of the energy yield a value smaller than the FCI result, and the number of occurrences is larger for the VV than the NV$^-$ center. These incorrect results are a manifestation of the so called unphysical state problem~\cite{sawaya2016error, mcardle2019error}, where in some cases quantum circuits yield states with an incorrect number of electrons causing an inaccurate evaluation of the energy. The unphysical state problem was first pointed out by Sawaya et al.~\cite{sawaya2016error} using simple noise models to study second row dimers, and several methods to correct for unphysical states were recently proposed, including error detection using ancilla qubits~\cite{mcardle2019error}.

We found that the apparent violation of the variational principle does pose a serious problem to the applicability of the VQE algorithm to the calculations of the electronic properties of spin defects in materials. The severity of this problem also depends on the chosen Fermion-to-qubit transformation, which in turn determines the extent to which the qubit Hilbert space differs from the configuration state space spanned by the Fermionic Hamiltonian~\cite{shee2021qubit}.

We adopt a post-selection method to enforce the validity of the variational principle during the VQE optimization. After using the Fermion-to-qubit transformation, the energy is evaluated on the quantum hardware as the weighted sum of the expectation values of Pauli correlators, i.e., $E=\sum_i g_i \langle\hat{P}_i\rangle$. The expectation values of all Pauli correlators were obtained by measuring 8192 times $N_c$ independent circuits, where $N_c$ is the number of groups of Pauli correlators that contain commuting operators. The number of electrons may be determined simultaneously from the group of operators that only contains diagonal Pauli correlators (those with only $I$ and $Z$ gates)~\cite{elfving2021simulating}. In our calculations we discard all measurement outcomes that do not conserve the number of electrons. Interestingly, the weights $g_i$ of diagonal Pauli correlators lead to the dominant contribution to the energy. The orange curves in Fig.~\ref{fig:results} A and B show the convergence of the VQE algorithm when energy measurements are obtained with the post-selection method, and Fig.~\ref{fig:results} C shows the convergence of the variational parameter $\theta$ with a relative error of less than $2\%$. Upon enforcement of the post-selection rule all measured energies turn out to be higher than the reference value for the ground state. Interestingly, the same post-selection method has also been adopted in the calculation of the total energy of LiH~\cite{elfving2021simulating}, yielding a notable improvement in the accuracy of energy measurements, with a small overall error of about 1 kcal/mol.

\begin{figure}[t!]
\includegraphics[width=0.7\textwidth]{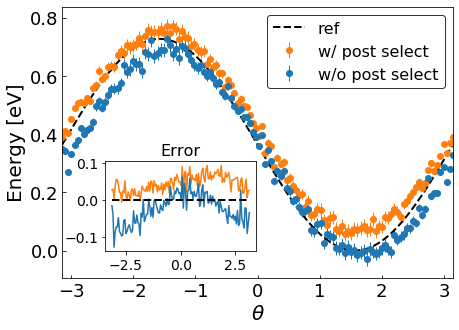}
\caption{Energy variation of the ground state of the $\text{NV}^-$ center in diamond as a function of the parameter $\theta$ (see Eq.~\ref{eq:Ansatz} in text) using \textit{ibmq\_casablanca}. We show results with (orange) and without post-selection of states (blue). The straight black line (ref) indicates the energy obtained with a noiseless simulator. Inset: difference between the energy evaluated on quantum and classical hardware. \label{fig:casablanca_scan}}
\end{figure}

In Fig.~\ref{fig:casablanca_scan} we further analyze the effects of noise on the results by scanning the total energy as a function of $\theta$: $E(\theta = \frac{\pi}{2})$ is the energy of the ground state and $E(\theta = -\frac{\pi}{2})$ that of an excited state. In the region close to the minimum (ground state) the values obtained with post selection (orange curve) are higher than the noiseless reference values (dashed black curve): indeed the post-selection process removes states that, due to the presence of noise, do not correspond to the correct number of electrons; hence, in virtue of the variational principle, one obtains an energy higher than the reference value. However in the region close to $\theta = - \frac{\pi}{2}$ (excited state) where the variational principle does not hold, even after post-selection one may obtain errors of different signs. Therefore a spurious cancellation of errors may occur and the overall error on the energy for $\theta < 0$ appears to be smaller than in proximity of the ground state ($\frac{\pi}{2})$).

To improve the accuracy of energy measurements, numerous mitigation schemes have been developed, including the quasi-probability method~\cite{temme2017error, endo2018practical}, individual error reduction~\cite{otten2019accounting}, and learning-based error mitigation~\cite{strikis2020learning,czarnik2020error,zlokapa2020deep,rogers2021error}. These methods require complete information of the noise channel or a large number of quantum measurements. We note that quantum noise is usually described using the language of open quantum systems and Kraus operators, and exact forms of the these operators~\cite{endo2021hybrid} are required in order to obtain a complete information of the noise channel, which is a difficult task for realistic hardware architectures. Here, we adopt the zero-noise extrapolation (ZNE) method, which is straightforward to implement and does not require additional qubits. The essence of the method is the expansion of the expectation value of an observable, for instance the energy $E$,  as a power series of noise around its zero-noise value $E^*$~\cite{kandala2019, endo2021hybrid}:
\begin{equation}
    E(\lambda) = E^* + \sum_{k=1}^n a_k \lambda^k + O(\lambda^{n+1}). \label{eq:noise_expansion}
\end{equation}
Here $\lambda$ is an appropriately small ($\lambda \ll 1$)  noise parameter, and the coefficients in the expansion $a_k$ depend on specific details of the noise (i.e., on the unknown form of the Kraus operators describing the specific noise channels). The basic idea of ZNE is to amplify the noise of the circuit to various controllable levels and obtain the zero noise limit by extrapolation. Error mitigation via ZNE has been explored in several pioneering papers~\cite{dumitrescu2018cloud, kandala2019, fauseweh2021digital} investigating small molecules and Ising Hamiltonians. Popular ways to artificially boost the error include identity insertions~\cite{he2020zero}, unitary folding~\cite{giurgica2020digital}, Pauli twirling~\cite{wallman2016noise,li2017efficient} and re-scaling of the Hamiltonian~\cite{temme2017error,kandala2019error,tomkins2020noise}.

Here, we propose a simple technique to boost the error for the ZNE, which we call exponential block replication. The method is applicable to all cases where the UCC ansatz~\cite{lee2018generalized,ryabinkin2018qubit,bauman2019downfolding,metcalf2020resource,fedorov2021unitary, liu2021unitary} is used. As discussed in Sec.~\ref{VQE}, UCC ansätze are typically implemented using the approximation $e^{\hat{A}+\hat{B}} \approx e^{\hat{A}} e^{\hat{B}}$. In this case multiple consecutive applications of $n$ exponential blocks lead to the following expression: $e^{\hat{A}+\hat{B}} \approx e^{\hat{A}} e^{\hat{B}} = \left(e^{\frac{\hat{A}}{n}} \right)^n \left(e^{\frac{\hat{B}}{n}} \right)^n$. This expression is different from that obtained with a $n$-step first-order Trotter decomposition, where  $e^{\hat{A}+\hat{B}} \approx \left(e^{\frac{\hat{A}}{n}} e^{\frac{\hat{B}}{n}}\right)^n$. By adopting this procedure we successfully increase the overall depth of the circuit by integer multiples of the original block and we artificially amplify the noise level without modifying the expression of the many-body wave-function or affecting the Trotter error.

\begin{figure}[h!]
    \centering
    \includegraphics[width=0.7\textwidth]{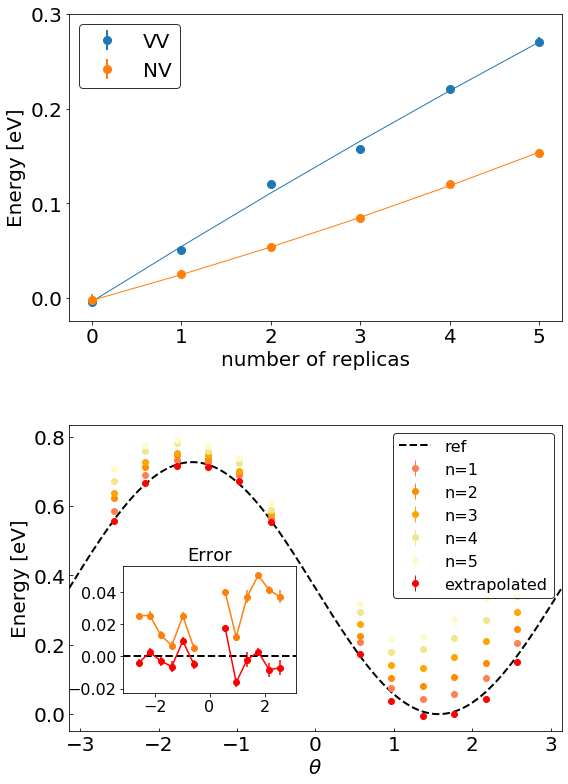}
    \caption{The upper panel shows the ground state energy of the NV$^-$ center and the VV in 4H-SiC as a function of number of replicas used in the zero-noise extrapolation (see text). The reference, noiseless result has been set at 0. The lower panel shows the total energy of the NV$^-$ center as a function of the parameter $\theta$ (Eq.~\ref{eq:Ansatz} in text). The different colors correspond to using $n=[1,2,3,4,5]$ replicas of exponential blocks and the red dots denote the linearly extrapolated energy ($n\to0$). In the inset we show the difference between noisy energies ($n=1$) or extrapolated values ($n\to0$) and the noiseless reference energy. Both panels are obtained using \textit{ibmq\_casablanca}.}
    \label{fig:extrapolation}
\end{figure}

Our extrapolation procedure is shown in Fig.~\ref{fig:extrapolation}. We considered $n=[1,2,3,4,5]$, and for each value of $n$ we repeated the 8912 measurements of each circuit 50 times (for a total of 50$\times$8912 = 445,600 measurements), in order to improve the stability of the extrapolation procedure. Repeating the measurement of each circuit 25 times instead of 50 leads to a small difference of $\sim$0.012 meV in the computed energy. The difference between the ground state energy obtained with 50 repetitions and the reference value obtained on a quantum simulator is $\sim$0.002 eV for the NV$^-$ center and $\sim$0.005 eV for the VV in SiC, one order of magnitude smaller than in the case where ZNE is not applied.

Fig.~\ref{fig:extrapolation} also shows the results of ZNE for the energy of the NV$^-$ center as a function of $\theta$. The extrapolation is performed considering an increasing number of replicas ($n$), and averaging over 24 repetitions of 8912 measurements for each replica. The choice of 24 instead of 50 repetitions is a compromise between efficiency and accuracy: scanning $E(\theta)$ (i.e., computing 12 values of the energy with a 50 time repetition rate) would require a time to completion during which we could not insure a constant noise level on the available hardware. Since we showed earlier that 25 repetitions yielded an acceptable result for the ground state, we used a similar number for the calculation of the function $E(\theta)$. The inset of Fig.~\ref{fig:extrapolation} shows the difference between noisy results ($n=1$) and extrapolated values ($n\to0$) and the noiseless reference. Overall we see that results obtained with ZNE are in closer agreement with the reference results. The errors of extrapolation in Fig.~\ref{fig:extrapolation} is computed using the method discussed in Appendix~\ref{SI: Error}.

\subsection{Calculation of the excited states using a quantum computer} \label{QSE}

We computed excited states of the NV$^-$ and VV centers using the QSE algorithm. To avoid propagating the errors introduced by VQE, we used the exact energy of the ${}^3A_2$ state with $m_s=0$ as ground state energy.

We constructed a quantum subspace that is identical to the configuration state space, so the dimension of the QSE matrices is the same as that of their classical FCI counterpart. The QSE matrix is built by evaluating, on the quantum hardware, the expectation values of all Pauli correlators; in addition we adopted the post-selection and the ZNE techniques described above. To perform ZNE we used a linear extrapolation for the off-diagonal elements of the QSE matrix and we computed diagonal elements with linear, quadratic, or exponential extrapolations. The number of repetition of the same circuit with 8192 measurements was reduced from 50 to 10 for the  VV due to the computational cost. The QSE matrix was diagonalized on a classical computer.

The results are summarized in Table \ref{tab:table1}. The accuracy of the energy of non-degenerate excitations is in general improved when using the ZNE, with the only exception being the ${}^{3}A_2 \leftrightarrow {}^{3}E$ transition of NV$^-$ in diamond obtained with a quadratic extrapolation scheme. We note that overall different choices of extrapolation functions lead to similar results, and hence linear extrapolation is a desirable choice, since a smaller number of parameters is expected to lead to a more stable fit. Unfortunately, we find that the degeneracy of states is spuriously lifted on the quantum hardware due to the presence of noise, even after applying the ZNE, showing that it is not possible to resolve energy differences smaller than the standard deviation ($\sigma$) associated to the mean of our  measurements. In our calculations with 50 repetitions $\sigma \simeq$ 3 meV.

\begin{table}
\caption{\label{tab:table1} Excitation energies [eV] of the NV$^-$ and VV centers calculated using the quantum subspace expansion (QSE) method. The first column  shows  transition between states labeled using the representation of the point group $C_{3v}$, following~\cite{ma2020a}. The second column shows results obtained with a noiseless simulator that are identical to those of classical full configuration interaction (FCI) calculations on a classical computer. The 3rd to 6th columns display results obtained using QSE on the quantum hardware, and using post-selection of states with different extrapolation strategies and the zero noise extrapolation technique.}
\begin{ruledtabular}
\begin{tabular}{cccccc}
NV$^-$ center  & FCI/noiseless & No extrap. &  Linear\footnotemark[1] & Quadratic\footnotemark[2] & Exponential\footnotemark[3]\\ 
\hline
${}^{3}A_2 \leftrightarrow {}^{1}E$\;\footnotemark[4] & 0.512 & 0.470 & 0.511 & 0.508 & 0.509\\
${}^{1}E \leftrightarrow {}^{1}E$\;\footnotemark[5] & 0.000 & 0.076 & 0.074 & 0.084 & 0.080\\
${}^{3}A_2 \leftrightarrow {}^{1}A_1$ & 1.380 & 1.282 & 1.391 & 1.373 & 1.378\\
${}^{3}A_2 \leftrightarrow {}^{3}E$\;\;\footnotemark[4] & 2.008 & 1.964 & 1.989 & 1.946 & 1.974\\
${}^{3}E \leftrightarrow {}^{3}E$\;\;\footnotemark[5] & 0.000 & 0.177 & 0.119 & 0.059 & 0.091\\
\hline
VV in SiC & FCI/noiseless & No extrap. & Linear\footnotemark[1] & Quadratic\footnotemark[2] & Exponential\footnotemark[3]\\
\hline
${}^{3}A_2 \leftrightarrow {}^{1}E$\;\;\footnotemark[4] & 0.378 & 0.338 & 0.367 & 0.363 & 0.365\\
${}^{1}E \leftrightarrow {}^{1}E$\;\;\footnotemark[5] & 0.002 & 0.083 & 0.069 & 0.065 & 0.067\\
${}^{3}A_2 \leftrightarrow {}^{1}A_1$ & 1.228 & 1.141 & 1.212 & 1.207 & 1.202\\
${}^{3}A_2 \leftrightarrow {}^{3}E$\;\;\footnotemark[4] & 1.348 & 1.313 & 1.337 & 1.333 & 1.334\\
${}^{3}E \leftrightarrow {}^{3}E$\;\;\footnotemark[5] & 0.002 & 0.010 & 0.080 & 0.079 & 0.078
\end{tabular}
\end{ruledtabular}
\footnotetext[1]{Linear extrapolation of both diagonal and off-diagonal elements of the QSE matrix (Eq.~\ref{eq:noise_expansion} in the text).}
\footnotetext[2]{Quadratic extrapolation of the diagonal, and linear of the off-diagonal elements of the QSE matrix.}
\footnotetext[3]{Exponential extrapolation of the diagonal, and linear of the off-diagonal elements of the QSE matrix.}
\footnotetext[4]{Energy of the degenerate states is computed as the average of the two energies obtained on quantum hardware.}
\footnotetext[5]{Energy gap between two states which should be degenerate, due to the presence of noise.}

\end{table}

\clearpage

\section{Conclusions} \label{Conclusions}


In summary, we presented electronic structure calculations of strongly correlated ground states and, for the first time, of excited electronic states of point defects in semiconductors on a near-term quantum computer. We focused on two spin-defects, i.e., the NV$^-$ center in diamond and the VV in SiC. Our computational protocol includes first principles calculations of the electronic structure of a spin defect in a solid containing hundreds of atoms using hybrid DFT, followed by the use of the quantum defect embedding theory to define an effective Hamiltonian that represents electronic excitations localized within the point-defect. The Hamiltonian is then mapped into a qubit Hamiltonian which is solved using VQE and QSE hybrid quantum/classical algorithms to obtain the ground and excited many-body electronic states of the defect, respectively. We discussed the merits of these algorithms in the case of spin qubits; however, establishing which algorithms are better suited to obtain, in general, many-body energies of electronic states in solids on NISQ hardware remains an open area of research~\cite{bauer2020}. For example, recent papers have proposed methods to find the eigenstates of a Fermionic Hamiltonian that are not based on the variational principle and therefore do not require the definition of an ansatz or involve an optimization procedure~\cite{motta2020}. In particular, Ref~\cite{korol2021quantum} proposed an algorithm to prepare approximate ground states with shallow circuit  and just one parameter.

We also discussed two main problems arising from the presence of noise on quantum hardware: (i) the apparent violation of the variational principle in VQE calculations due to unphysical states arising when the number of electrons is not conserved, and (ii) the presence of persisting errors on energies obtained on quantum computers even after correcting for the presence of unphysical states. We successfully applied a post-selection method based on partial constraints on the number of electrons to correct for problem (i) and we proposed an error mitigation technique within the ZNE scheme to reduce the effect of quantum errors. The technique uses an exponential block repetition to boost the quantum error of UCC type ans\"atze in a controllable fashion. The error mitigation protocol adopted here has several advantages: (1) it is readily applicable without any knowledge of the source of hardware noise and without increasing the number of qubits; (2) it does not affect the scaling of the quantum algorithm, although it may affect the prefactor. However, the UCC type ans\"atze require the use of relatively deep circuits thus limiting the applicability of ZNE strategies with a large number of replicas. As a proof of principle of the strategies adopted here to solve useful materials problems, we obtained results with small active spaces and shallow circuits. Work is in progress to expand the applicability of the method to systems that require larger active spaces appropriate to investigate, for example, adsorbates on surfaces and ions or nanostructures in solution. Based on our work, we further envision a feedback loop, where quantum simulations of materials properties on a quantum device lead to the prediction of new materials and properties for the design of improved quantum computers, which will in turn result in enhanced property predictions and applications, therefore establishing a tight connection between quantum computations and materials' predictions.

\clearpage

\begin{acknowledgments}
We thank He Ma, Nan Sheng, Christian Vorwerk, Liang Jiang, and Martin Suchara for many fruitful discussions. We also thank the Qiskit Slack channel for generous help.  This work was supported by the computational materials science center MICCoM for the implementation and use of quantum embedding and by QNEXT hub for the development of quantum algorithms and deployment on quantum hardware. MICCoM is part of the Computational Materials Sciences Program funded by the U.S. Department of Energy, Office of Science, Basic Energy Sciences, Materials Sciences, and Engineering Division through Argonne National Laboratory, under contract number DE-AC02-06CH11357. QNEXT sis supported by the U.S. Department of Energy, Office of Science, National Quantum Information Science Research Centers. This research used resources of the Oak Ridge Leadership Computing Facility at the Oak Ridge National Laboratory, which is supported by the Office of Science of the U.S. Department of Energy under Contract No. DE-AC05-00OR22725, resources of the National Energy Research Scientific Computing Center (NERSC), a DOE Office of Science User Facility supported by the Office of Science of the US Department of Energy under Contract No. DE-AC02-05CH11231, and resources of the Argonne Leadership Computing Facility, which is a DOE Office of Science User Facility supported under Contract DE-AC02-06CH11357. We acknowledge the use of IBM Quantum services for this work. The views expressed are those of the authors, and do not reflect the official policy or position of IBM or the IBM Quantum team.

\end{acknowledgments}

\appendix

\section{Ansatz} \label{SI: Ansatz on trial states}

We discuss the case of the NV$-$ center in diamond as an example. We use a (4e, 3o) active space, as shown in Figure \ref{fig:classical_preparation}. After parity mapping and tapering off qubits, the ground state can be represented as:
\begin{equation}
    \ket{{}^3A_2, m_s=0} = \frac{1}{\sqrt{2}}\left(\ket{011001} + \ket{001011} \right) \xrightarrow{\text{tapering}} \frac{1}{\sqrt{2}}\left(\ket{1101} + \ket{0111} \right),
\end{equation}
where we denote the spin-up orbitals  with 0, 1, 2 and the spin-down orbitals with 3, 4, 5 in ascending order of energy. The zeroth and second qubit remains 1 after tapering, since the $a_1$ orbital is occupied. By selecting only spin-conserving excitations, we are left with one variational parameter $\theta_{15}^{24}$ ($\theta_{e_x \bar{e}_y}^{e_y \bar{e}_x}$) and the operator acting on the initial state may be written as:
\begin{equation}
    U(\theta_{15}^{24})\ket{1101} = e^{-\text{i}\frac{\theta_{15}^{24}}{4}(Y_1 X_3 - X_1 Y_3)}\ket{1101} = e^{\text{i}\frac{\theta_{15}^{24}}{2} X_1 Y_3} \ket{1101},
\end{equation}
where we have taken advantage of commutation rules~\cite{hempel2018}. This assumption leads to a simple ansatz circuit similar to that shown in Figure \ref{fig:ansatz_circuit}. When running on a quantum simulator, we obtain converged values within $10^{-9}$ eV.

\section{Error Extrapolation} \label{SI: Error}

The error in measuring $\braket{H}$ has two components: a systematic and a stochastic~\cite{peruzzo2014} component. In our extrapolation, we estimate the standard deviation of $\braket{H}$ at each number of circuit replicas $n$. In addition, we estimate the error present in the zero noise limit. To do so, we conducted 50 repetitions  of each measurement of $\braket{H(n)}$ and computed an average value $\overline{E}_n$ and distribution $\sigma_E$.

Assuming $\overline{E}_n$ as a function of the number of replicas $n$  can be approximated by a polynomial function, $\overline{E}$ may be used for both fitting and extrapolation procedures. Importantly, the error associated to $\overline{E}$ is $\sigma_{\overline{E}} = \sigma_{E} / \sqrt{N}$, where here $N$ is the number of independent (repeated) measurements. We call the error on  $\overline{E}$ in the zero noise limit: $\overline{E}_{obj}\pm {\sigma}_{obj}$.

We first fit the data sets \{${n_i, i=1\cdots m}$\} and \{${\overline{E_i}, i=1\cdots m}$\} with a polynomial function. Using a least squares fit, the function to be determined is parametrized as:
\begin{align}
    \overline{E} = \sum_{i=0}^{p}\alpha_i n^i,
\end{align}
and the parameters can be obtained by solving the following linear system $L \alpha = K$, where 
\begin{equation}
   L = 
   \begin{bmatrix}
   m & \sum_{i=1}^m n_i & \dots & \sum_{i=1}^m n_i^p \\
   \sum_{i=1}^m n_i & \sum_{i=1}^m n_i^2 & \dots & \sum_{i=1}^m n_i^{p+1} \\
   \vdots & \vdots & \ddots & \dots\\
   \sum_{i=1}^m n_i^p & \sum_{i=1}^m n_i^{p+1} & \dots & \sum_{i=1}^m n_i^{2p}
   \end{bmatrix},\;
   \alpha = 
   \begin{bmatrix}
   \alpha_0 \\
   \alpha_1 \\
   \vdots \\
   \alpha_p
   \end{bmatrix},\;
   K =
   \begin{bmatrix}
   \sum_{i=1}^m \overline{E_i} \\
   \sum_{i=1}^m n_i \overline{E_i} \\
   \vdots \\
   \sum_{i=1}^m n_i^p \overline{E_i}
   \end{bmatrix}
\end{equation}
The result briefly reads $\alpha = L^{-1} K$, where each $\alpha_i$ is a linear combination of \{${\overline{E}_i, i=1 \cdots m}$\}. $\alpha$ can be expressed as: $\alpha = A E$, where $A$ is a $(p+1) \times m$ matrix with entries $\{A_{ij}\}$, and $E$ is a $m\times$1 matrix with entries $\{\overline{E}_i\}$. The standard deviation of $\theta$, which is a $(p+1)$ dimensional vector, satisfies the following equation:
\begin{align}
    \sigma_{\alpha} \odot \sigma_{\alpha} = (A \odot A) (\sigma_{\delta{d}} \odot \sigma_{\delta{d}}) .
\end{align}
Here $\sigma_{\overline{E}}$ is not a scalar but a $m \times 1$ matrix with entries $\{\sigma_{\overline{E}_i}\}$. $\odot$ is the Hadamard product. The quantity $\overline{E}_{obj}$ is given by $\alpha_0$, and we obtain:
\begin{align}
    \alpha_0 \pm \sigma_{\alpha_{0}}.
\end{align}

\section{Description of quantum hardware} \label{SI: Hardware}

\begin{figure}[h!]
    \centering
    \includegraphics[width=0.3\textwidth]{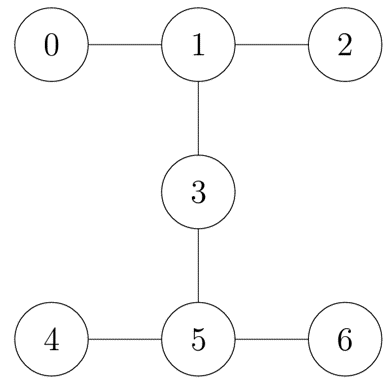}
    \caption{The 7-qubit \textit{ibmq\_casablanca} device layout.}
    \label{fig:hardware_layout}
\end{figure}

We used the \textit{ibmq\_casablanca} processor running Qiskit v0.24.1~\cite{aleksandrowicz2019qiskit}. The layout of this quantum device is shown in Fig.~\ref{fig:hardware_layout}, where the coupling qubits are connected by a solid line. Table~\ref{tab:table2} summarizes the coherence time, single-qubit, two-qubit and readout error rates of the device when the measurements reported in the main text were executed. In choosing the qubit device mappings shown in the main text, preference was given to the qubit pairs with relatively small CNOT error rates. In our experiments, CNOT gates are executed between Q1 and Q3 on the device.

\begin{table}
\caption{\label{tab:table2} Calibration data for the \textit{ibmq\_casablanca} device. $T_1, T_2$ denotes the coherence time of the qubits. The 4th and 5th column display the single qubit $X$ rotation and the readout error rate of each qubit. And the last column displays the two-qubit CNOT error rate between coupling qubit pairs.}
\begin{ruledtabular}
\begin{tabular}{c|cc|cc||cc}
Qubit & $T_1\;(\mu s)$ & $T_2\;(\mu s)$ & $X\;(10^{-4})$ & Readout $(10^{-2})$ & Qubit pair & CX  $(10^{-2})$\\ 
\hline
Q0 & 51 & 33 & 4.0 & 4.3 & [Q0, Q1] & 1.484\\
Q1 & 110 & 105 & 2.0 & 2.2 & [Q1, Q2] & 0.957\\
Q2 & 114 & 149 & 3.5 & 2.9 & [Q1, Q3] & 0.671\\
Q3 & 102 & 201 & 3.9 & 2.0 & [Q3, Q5] & 1.168\\
Q4 & 107 & 75 & 2.4 & 3.1 & [Q4, Q5] & 1.091\\
Q5 & 37 & 72 & 4.2 & 1.2 & [Q5, Q6] & 0.948\\
Q6 & 107 & 188 & 6.4 & 2.4 &  &
\end{tabular}
\end{ruledtabular}
\end{table}

In addition to the errors discussed in the main text, the state preparation and measurement processes are noisy as well and these errors are usually denoted as state preparation and measurement errors (SPAM). To reduce the effect of noise due to final measurement errors, Qiskit recommends a measurement error mitigation approach~\cite{maciejewski2020mitigation}. To do so, we adopted a recommended calibration procedure on the chosen basis states before each measurement of the energies, in order to characterize the device noise. After each  calibration measurement, we arranged the results in a $2^N \times 2^N$ matrix $C$,
\begin{equation}
    C_{ij} = p(i,j),
\end{equation}
where $p(i,j)$ is the probability of preparing state $i$ and measuring state $j$. If the single gate errors are small compared to the readout noise this matrix perfectly maps the ideal, expected  results onto the measured results using a classical post processing of the statistics
\begin{equation}
    p_{\text{exp}} = C p_{\text{ideal}}.
\end{equation}
Thus to obtain the ideal results we applied the inverse of $C$ onto our measurements. All the manipulations described above are implemented in the Qiskit package~\cite{Qiskit-Textbook} and used in our study.

\bibliography{bibliography}

\end{document}